\documentclass[journal]{IEEEtran}

\usepackage{cite}

\ifCLASSINFOpdf
  \usepackage[pdftex]{graphicx}
  
\else
  
\fi

\usepackage{amsmath}

\usepackage{algorithmic}

\usepackage{array}

\ifCLASSOPTIONcompsoc
  \usepackage[caption=false,font=normalsize,labelfont=sf,textfont=sf]{subfig}
\else
  \usepackage[caption=false,font=footnotesize]{subfig}
\fi

\usepackage{fixltx2e}

\usepackage{dblfloatfix}

\ifCLASSOPTIONcaptionsoff
  \usepackage[nomarkers]{endfloat}
 \let\MYoriglatexcaption\caption
 \renewcommand{\caption}[2][\relax]{\MYoriglatexcaption[#2]{#2}}
\fi

\usepackage{url}

\usepackage{multirow}
\newcolumntype{P}[1]{>{\centering\arraybackslash}p{#1}}
\usepackage{caption}
\usepackage{hhline}

\usepackage[hidelinks]{hyperref}

\begin{document}

\title{Light Field Image Quality Assessment with Auxiliary Learning based on Depthwise and Anglewise Separable Convolutions}

\author{Qiang~Qu,
        Xiaoming~Chen,
        Vera~Chung,~\IEEEmembership{Member,~IEEE,}
        and~Zhibo~Chen,~\IEEEmembership{Senior~Member,~IEEE}% <-this % stops a space
% \thanks{M. Shell was with the Department
% of Electrical and Computer Engineering, Georgia Institute of Technology, Atlanta,
% GA, 30332 USA e-mail: (see http://www.michaelshell.org/contact.html).}% <-this % stops a space
% \thanks{J. Doe and J. Doe are with Anonymous University.}% <-this % stops a space
\thanks{Date of publication 21 December 2021}}

\maketitle

\begin{abstract}

In multimedia broadcasting, no-reference image quality assessment (NR-IQA) is used to indicate the user-perceived quality of experience (QoE) and to support intelligent data transmission while optimizing user experience. This paper proposes an improved no-reference light field image quality assessment (NR-LFIQA) metric for future immersive media broadcasting services. First, we extend the concept of depthwise separable convolution (DSC) to the spatial domain of light field image (LFI) and introduce ``light field depthwise separable convolution (LF-DSC)'', which can extract the LFI’s spatial features efficiently. Second, we further theoretically extend the LF-DSC to the angular space of LFI and introduce the novel concept of ``light field anglewise separable convolution (LF-ASC)'', which is capable of extracting both the spatial and angular features for comprehensive quality assessment with low complexity. Third, we define the spatial and angular feature estimations as auxiliary tasks in aiding the primary NR-LFIQA task by providing spatial and angular quality features as hints. To the best of our knowledge, this work is the first exploration of deep auxiliary learning with spatial-angular hints on NR-LFIQA. Experiments were conducted in mainstream LFI datasets such as Win5-LID and SMART with comparisons to the mainstream full reference IQA metrics as well as the state-of-the-art NR-LFIQA methods. The experimental results show that the proposed metric yields overall 42.86\% and 45.95\% smaller prediction errors than the second-best benchmarking metric in Win5-LID and SMART, respectively. In some challenging cases with particular distortion types, the proposed metric can reduce the errors significantly by more than 60\%.
\end{abstract}

\begin{IEEEkeywords}
No-reference Image Quality Assessment, Quality of Experience, Light Field Image, Immersive Media, Auxiliary Learning, Deep Learning.
\end{IEEEkeywords}

\markboth{IEEE TRANSACTIONS ON BROADCASTING,~Vol.~67, No. 4, 2021}%
{Shell \MakeLowercase{\textit{et al.}}: A Sample Article Using IEEEtran.cls for IEEE Journals}

% \fi
\IEEEpeerreviewmaketitle

\section{Introduction}
\label{sec:introduction}
\IEEEPARstart
{T}{he} immersive media experience is presently undergoing a transition from three degrees of freedom (3 DoF) to six degrees of freedom (6 DoF), which provides a greater sense of immersion and enhanced interactivity \cite{song20166}. Apart from the ability to change viewing directions through rotation (i.e., pitch, yaw, and roll in 3 DoF), 6 DoF offers three additional degrees of freedom (i.e., surge, heave, and sway). Thus, users may navigate in the virtual world by changing not only the viewing directions but also the viewing positions. With recent advancements in VR and 5G networking technologies, immersive media with 6 DoF is anticipated to be dominant in future broadcasting services \cite{paudyal2019reduced}. Indeed, leading technology companies have already been engaged in this race for some time. For instance, Google developed a system in 2020 for recording and producing 6 DoF immersive light field videos \cite{broxton2020immersive}. Apple has filed a series of patents for 6 DoF light field imaging for next-generation mobile devices \cite{AppleLFC}. Additionally, Sony has announced an eye-sensing light field display capable of providing a 6 DoF viewing experience in late 2020 \cite{SonyLFD}.

Light field photography is a cornerstone of 6 DoF immersive media. In contrast to conventional two-dimensional imaging, it makes use of additional angular information by collecting light beams from multiple directions. In practice, light field images (LFIs) are often captured using an array of cameras rather than a single camera. As a result, it adds a new angular domain in addition to the spatial domain inherited from 2D photography. Due to this property, LFI is capable of providing a 6 DoF experience in future immersive media applications. However, the advent of LFI imposes new research challenges related to LFI processing. Particularly, no-reference light field image quality assessment (NR-LFIQA) is of great significance since it can quantify the user's quality of experience (QoE) and therefore assist intelligent LFI transmission. Due to the unique structural features of LFI, however, current 2D IQA techniques are unable to accurately predict the user-perceived quality. Although there are some recently proposed NR-LFIQA methods such as BELIEF \cite{shi2019belif} and NR-LFQA \cite{shi2019no}, their performance is limited as they are fundamentally designed based on 2D IQA methodologies such as naturalness statistics and structural similarity (SSIM) \cite{wang2004image}. As a result, they gain sub-optimal performance and predict inaccurately on certain distortion types such as EPICNN \cite{wu2017light} and USCD \cite{kalantari2016learning} (See Section \ref{sec:experiments} for details).

To overcome the drawbacks of the aforementioned methods, we propose in this paper a more precise NR-LFIQA metric using a new framework termed ``auxiliary learning with angular and spatial features via deep anglewise and depthwise separable convolutions (ALAS-DADS)''. We believe that the superiority of the proposed metric is due to the following features. First, since the LFI processing usually suffers from the curse of dimensionality, we propose the light field depthwise separable convolution (LF-DSC)\cite{howard2017mobilenets} as a low-complexity feature extractor for evaluating the spatial quality of an LFI. Second, to supplement the angular quality information missed by LF-DSC, we further propose the novel light field anglewise separable convolution (LF-ASC) to extract both spatial and angular features for comprehensive quality assessment. Third, we develop an auxiliary learning scheme with two sub-tasks: spatial quality feature estimation and angular quality feature estimation. This scheme leverages both angular and spatial quality features as hints to assists the primary NR-LFIQA task.

The main contributions of this paper are summarised as follows.
\begin{itemize}
    \item We theoretically extend the traditional depthwise separable convolution (DSC) to LFI space as LF-DSC. Moreover, we propose the novel LF-ASC for comprehensive LFI feature extraction. 
    \item Based on LF-DSC and LF-ASC, we propose the first deep auxiliary learning-based model with spatial-angular hints on no-reference light field image quality assessment (NR-LFIQA), which has been verified to outperform the mainstream full reference IQA metrics as well as the state-of-the-art NR-LFIQA methods significantly.
    \item We not only show the effectiveness of the proposed LF-DSC, LF-ASC, and the auxiliary learning model in NR-LFIQA, but also shed light on their potential use in other LFI processing tasks such as LFI super-resolution, depth estimation, and so on, paving the way for future LFI research.
\end{itemize}

 The remaining of this paper is structured as follows. 
 In Section \ref{sec:relatedworks}, we review some related works covering light field imaging, IQA, NR-LFIQA, separable convolution, and auxiliary learning. 
 In Section \ref{sec:methods}, we illustrate and describe the overall structure, LF-DSC, LF-ASC, and auxiliary learning components of the proposed metric. We have conducted extensive experiments and analyzed the experimental results in Section \ref{sec:experiments}. In Section \ref{sec:conclusion}, we conclude our work and discuss future work.
 
\section{Related Works}
\label{sec:relatedworks}
\subsection{Light Field Imaging}

% \begin{figure}[t!]
% \includegraphics[width=\linewidth]{images/plenoptic_camera.png}
% \caption{Demonstration of the Plenoptic Photography}
% \label{fig:plen_photo}
% \end{figure}
Unlike traditional photography, which records the 2D projection of light rays, an LFI describes light rays from multiple directions. This technology allows capturing richer visual information due to the higher dimensional representation of the data \cite{ihrke2016principles}. The theoretical root of LFI is interpreted by a plenoptic function, which is usually denoted as $l_{\lambda}(x,y,z,\theta,\phi, \lambda, t)$ where $l_{\lambda}[W/m^2/sr/nm/s]$ represents spectral radiance per unit time, with $(x,y,z)$ for the spatial position, $(\theta,\phi)$ for light ray directions, $\lambda$ for wavelength, and $t$ is a temporal instance \cite{ihrke2016principles}. With a series of background assumptions, each point in a light field can be denoted as a 4D coordinate $(u,v,x,y)$ where $(x,y)$ represents the spatial position and $(u,v)$ indicates the angular position \cite{ihrke2016principles}. From the digital imaging perspective, each pixel in an LFI can be located with two 2D coordinates: an angular coordinate identifying the located subview in the LFI collection and a spatial coordinate indicating the exact pixel position in that subview similar to a traditional 2D image.

\subsection{Image Quality Assessment}
\subsubsection{Overview on IQA}
\label{subsec:iqa}
According to the availability of original reference data, objective IQA algorithms can be categorised as full-reference IQA (FR-IQA), reduced-reference IQA (RR-IQA), or no-reference IQA (NR-IQA). The FR-IQA metrics assess image quality categorically by comparing the distorted images to the reference images. The RR-IQA metrics only use partial information of the reference image in the IQA process. The NR-IQA metrics, in contrast to the previous two, evaluate the image quality without using original reference images, which is more practical and relevant in real-world situations.

2D IQA is a well-studied problem with numerous 2D FR-IQA metrics proposed such as SSIM, \cite{wang2004image}, multi-scale SSIM \cite{wang2003multiscale}, FSIM \cite{zhang2011fsim}, IWSSIM \cite{wang2010information}, visual saliency-induced index (VSI) \cite{zhang2014vsi}, gradient magnitude similarity deviation (GMSD) \cite{xue2013gradient}, efficient Minkowski distance-based metric \cite{zia2018distort}, multiple pseudo reference images (MPRIs) similarity index \cite{min2018distort}, and multiscale edge attention (MSEA) similarity index \cite{Yang2020screen}. Furthermore, some 2D RR-IQA indicators are also introduced. For example, the DIIVINE model measures the local magnitude and phase statistics of natural scenes to evaluate the image quality \cite{moorthy2011blind}. BLIINDS-II adopts the generalized Gaussian density function with block discrete cosine transform (DCT) coefficients \cite{moorthy2011blind} to extract the quality features. As one of the most popular NR-IQA metrics, BRISQUE uses the scene statistics of locally normalized luminance coefficients to quantify possible losses of “naturalness” owing to the presence of distortions \cite{mittal2012no}.

In comparison to the FR-IQA and RR-IQA techniques, the NR-IQA approaches have more technical challenges since they are entirely based on the distorted image that is being assessed to make a prediction. NR-IQA methods usually detect and extract distortion-specific features that lead to visual quality degradation \cite{ferzli2009no, suthaharan2009no, varadarajan2008improved}. However, these methods are incapable of recognising the inherent high-dimensional characteristics of LFI for the purpose of evaluating its visual quality.

\subsubsection{State-of-the-art Models for NR-LFIQA}
\label{subsec:state_of_art}
Nevertheless, there are few LFI-specific algorithms in the context of NR-LFIQA. Ideally, a comprehensive LFI-specific algorithm should consider both spatial quality and angular consistency. For example, recent work on NR-LFIQA includes BELIEF \cite{shi2019belif}, Tensor-NLFQ \cite{zhou2020tensor}, LGF-LFC \cite{tian2020light}, and NR-LFQA \cite{shi2019no}. Particularly, NR-LFQA proposed gradient direction distribution (GDD) to measure the deterioration of angular consistency. It involves the statistical analysis of horizontal and vertical gradient direction maps. According to the experimental results, the NR-LFQA outperforms the others. For this reason, we have chosen NR-LFQA as one of the benchmarking models (see Section \ref{sec:experiments}).

\begin{figure*}[t!]
\centering
\includegraphics[width=\textwidth]{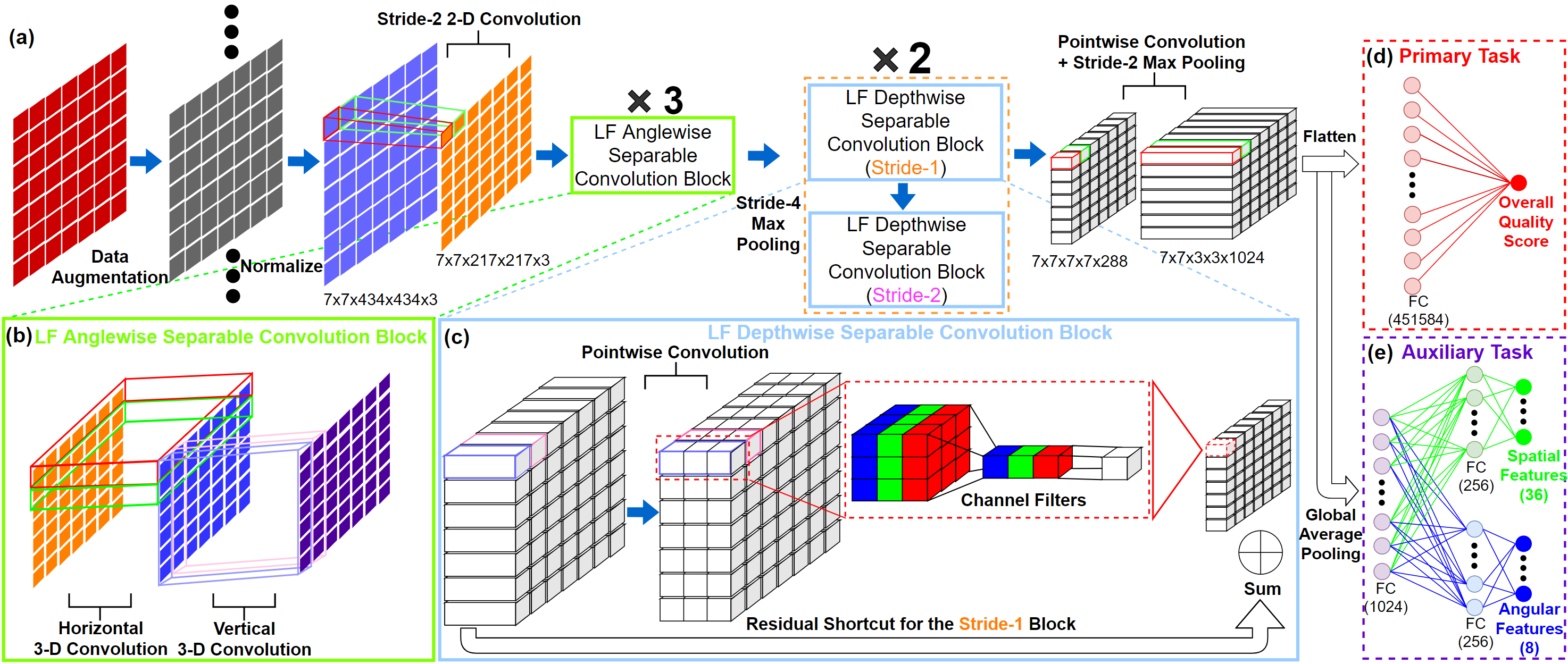}
\caption{The Architecture of the ALAS-DADS Framework}
\label{fig:model_archit}
\end{figure*}

\subsection{Separable Convolution and Auxiliary Learning}
\subsubsection{Separable Convolution}
\label{subsec:separable_convolution}
Separable convolution is a type of convolutional filter in convolutional neural networks (CNN). CNN is known to perform better with deeper network depth \cite{abdi2016multi}. However, increasing the number of stacked layers (depth) can be as challenging as the vanishing/exploding gradient problem appears during the learning process \cite{abdi2016multi}. Deep residual networks circumvent this problem by using identity skip-connections, which facilitate the training of much deeper networks \cite{targ2016resnet}.

Moreover, high computational intensity is always a problem for deep learning. Howard \textit{et al.} \cite{howard2017mobilenets} addressed this issue by introducing the 2D DSC in MobileNet. Unlike traditional 2D convolution, the DSC adopts an intermediate channel layer filter containing channel filters. The channel filters then extract each channel's features separately instead of all channels together like the regular 2D filters. The output of the channel filters then goes through a $1\times 1$ convolution that combines all channels' features. The separation between the color feature extraction and combination reinforces the feature extraction process and reduces the computing cost by a polynomial degree. Therefore, we believe that it is worth to explore how to extend DSC to LFI processing to reduce the computational complexity.

\subsubsection{Auxiliary Learning}
\label{subsec:rw_aux_learn}
According to Ruder \cite{ruder2017overview}, the introduction of auxiliary learning could be effective to achieve better performance if the associated features are used as hints for a primary task. Several examples of this approach have proved the feasibility of analogously in the field of natural language processing \cite{ruder2017overview}. Similar to the idea of ``hints'', auxiliary learning can also be used to force the learning to focus on the attention in the areas of interest that might be ignored in the regular training process \cite{ruder2017overview}.
There are several attempts to apply multi-task learning in NR-IQA. For example, Xu \textit{et al.} \cite{xu2016multi} proposed a multi-task learning framework to train multiple distortion type-specialised IQA models together. As an enhancement of Xu's model, instead of manually configuring models for each distortion type, Ma \textit{et al.} \cite{ma2017end} developed a multi-task deep learning model, MEON, which consists of a distortion identification network and a quality prediction network. Yan \textit{et al.} \cite{yan2019naturalness} introduced the natural statistics feature estimation sub-task to provide the learning model with an intuition for image naturalness.

\subsection{Summary of Previous Works}% Some Insights on Previous Works
In summary, although many IQA metrics have been proposed for the NR-LFIQA in last few decades, the majority of them are not intended for LFI and thus they often ignore to take advantage of the LFI's angular dimension. Several new NR-LFIQA metrics, including BELIEF and NF-LFQA, have yielded generally acceptable results. They are, however, inept in dealing with some forms of severe distortions such as EPICNN, resulting in instability in IQA prediction with large variations. As stated in Section \ref{subsec:separable_convolution}, DSC has been shown to be a high-performance and computationally efficient feature extractor even when the input dimension is limited to three dimensions (i.e., the pixel coordinates and a color channel). Thus, it is interesting to investigate the adaptation of DSC from conventional 2D images to LFI in order to substantially decrease the computational cost associated with the high dimensionality. Additionally, the idea of separable convolution may potentially be extended to the angular dimension in the context of LFI for further improvement in feature extraction. Moreover, as mentioned in Section \ref{subsec:rw_aux_learn}, the inclusion of auxiliary tasks may assist the model in mastering the primary task, which is NR-LFIQA in our case. Based on the above observations and analysis, we are inspired to present an auxiliary learning based model with effective LF-ASC and LF-DSC feature extractors, which is introduced in next section.

\section{Methodology}
\label{sec:methods}
The general structure of the proposed ALAS-DADS framework is explained in Section \ref{subsec:struct} and shown in Fig. \ref{fig:model_archit}. It comprises the LF-DSC module, the LF-ASC module, and the auxiliary learning based NR-LFIQA module. First, due to the fact that the original DSC was developed for 2D images and cannot be directly employed in LFI processing, we expand the concept of DSC to the spatial domain of LFI to form the LF-DSC module (see Section \ref{subsec:dep_con}). LF-DSC is a robust spatial feature extractor that is optimised for assessing spatial quality using a small number of parameters. Second, we theoretically expand the concept of LF-DSC to the angular space and introduce the novel concept of LF-ASC (see Section \ref{subsec:ang_con}). With significantly reduced computational cost, the LF-ASC can extract both spatial and angular features for complete NR-LFIQA. Third, we propose a novel auxiliary learning based module that exploits both available spatial and angular clues (described as sub-tasks) to aid the NR-LFIQA primary task (see Section \ref{subsec:aux_learn}). Finally, in Section \ref{subsec:dis}, we discuss how the proposed framework and techniques can be generalized in other LFI processing tasks.

%[Some content deleted form the above paragraph can be integrated in other parts of the manuscript.]  

\subsection{Model Structure}
\label{subsec:struct}
The ALAS-DADS framework is visualised in Fig. \ref{fig:model_archit} (a). The framework consists of two stride-1/stride-2 LF-DSC blocks and three LF-ASC blocks that are trained via auxiliary learning. The walkthrough of the proposed model can be summarised into the following process:
\begin{enumerate}
  \item Data augmentation is conducted to increase the dataset's size by eightfold. (see Section \ref{subsubsec:data_aug}).
  \item Normalization, i.e., for an LFI $n$:
  \begin{equation}\hat{p_n}(D)=\frac{p_n(D)-\mu(D)}{\sigma(D)+1}, \forall D=(u,v,x,y,c) \label{eq:norm}\end{equation} where $(u, v)$ identifies a subview and $(x, y, c)$ represents a pixel and its color channel. $\mu$ and $\sigma$ are mean and standard deviation of the corresponding pixel channel.
  \item A subview-wise stride-2 2D convolution is performed to reduce the spatial size.
  \item Three LF-ASCs are performed to extract angular as well as spatial features efficiently (see Section \ref{subsec:ang_con}).
  \item A stride-4 max pooling is applied to decrease the spatial size.
  \item Two stride-1/stride-2 LF-DSCs are performed for efficient spatial and angular feature extraction (see Section \ref{subsec:dep_con}).
  \item A pointwise convolution with the following stride-2 max pooling is applied to combine the channel features.
  \item The model outputs the NR-LFIQA score with auxiliary learning after fully connected neural networks (FC-NN) (see Section \ref{subsec:aux_learn}).
\end{enumerate}

To give more model details, TABLE \ref{tab:model_shapes} lists the filter shapes as well as the input size of each model layer. The input shape is $7\times 7\times 434\times 434\times 3$, and the model body will extract $7\times 7\times 3\times 3\times 1024$ features to feed in the prediction module. The module includes an FC-NN to predict the quality scores as the primary task, and two double 256-node layers to estimate spatial features and angular features, respectively.

\begin{table}
\renewcommand{\arraystretch}{1.2}
\caption{Model Details of ALAS-DADS}
\label{tab:model_shapes}
\centering
\setlength{\tabcolsep}{3pt}
\begin{tabular}{|p{0.15\textwidth}|p{0.15\textwidth}|p{0.15\textwidth}|} \hline
Type / Stride& Filter Shape& Input Size \\ \hline \hline
\multicolumn{3}{|c|}{Model Body}\\ \hline
2D Conv / s2 & $3\times 3\times 3\times 3$& $7\times 7\times 434\times 434\times 3$ \\ \hline
AW Conv / s1 & $7\times 7\times 4\times 4\times 3$ & $7\times 7\times 217\times 217\times 3$ \\ \hline
AW ConvBloc / s1 & $7\times 7\times 4\times 4\times 3$& $7\times 7\times 217\times 217\times 3$\\ \hline
AW ConvBloc / s1 & $7\times 7\times 4\times 4\times 3$& $7\times 7\times 217\times 217\times 3$\\ \hline
Max Pooling / s4 & N/A & $7\times 7\times 217\times 217\times 3$ \\ \hline
DW ConvBloc / s1 & $7\times 7\times 4\times 4\times 3$& $7\times 7\times 54\times 54\times 3$ \\ \hline
DW ConvBloc / s2 & $7\times 7\times 4\times 4\times 12$& $7\times 7\times 54\times 54\times 3$ \\ \hline
DW ConvBloc / s1 & $7\times 7\times 4\times 4\times 12$& $7\times 7\times 27\times 27\times 12$\\ \hline
DW ConvBloc / s2 & $7\times 7\times 4\times 4\times 48$ & $7\times 7\times 27\times 27\times 12$\\ \hline
DW ConvBloc / s1 & $7\times 7\times 4\times 4\times 48$& $7\times 7\times 14\times 14\times 48$\\ \hline
DW ConvBloc / s2 & $7\times 7\times 4\times 4\times 192$ & $7\times 7\times 14\times 14\times 48$\\ \hline
Pointwise Conv / s1 & $1\times 1\times 1\times 1024$ & $7\times 7\times 7\times 7\times 192$ \\ \hline
Max Pooling / s2 & N/A & $7\times 7\times 7\times 7\times 1024$ \\ \hline \hline
\multicolumn{3}{|c|}{Primary Task: IQA Score Prediction}\\ \hline
FC-NN & $451584$ & $7\times 7\times 3\times 3\times 1024$ \\ \hline
Output & N/A & $1$ \\ \hline \hline
\multicolumn{3}{|c|}{Auxiliary Task 1: Spatial Feature Estimation}\\ \hline
FC-NN & $256$ & $1024$ \\ \hline
FC-NN & $256$ & $1024$ \\ \hline
Output & N/A & $36$ \\ \hline \hline
\multicolumn{3}{|c|}{Auxiliary Task 2: Angular Feature Estimation}\\ \hline
FC-NN & $256$ & $1024$ \\ \hline
FC-NN & $256$ & $1024$ \\ \hline
Output & N/A & $8$ \\ \hline
\end{tabular}
\end{table}

\subsection{Depthwise Separable Convolutions for LFI}
\label{subsec:dep_con}
As stated in Section \ref{subsec:separable_convolution}, the DSC was initially developed for 2D images, which handle not only spatial dimensions but also depth dimensions (i.e., the number of channels). It converts a normal convolution to a depthwise convolution, then performs a $1\times 1$ convolution (i.e., pointwise convolution) to combine the outputs from the preceding layers. The primary difference between DSC and regular convolution is that the DSC separates the filtering features and combines the outputs into two layers instead of one step. Due to the fact that this factorization lowers the computing dimensions involved, it significantly reduces the computational complexity and the model size. To achieve these benefits, we expand the DSC into the spatial domain of the LFI to resolve the curse of dimensionality in LFI. Specifically, each subview in an LFI is treated as a 2D image that is fed into a depthwise convolution followed by a pointwise convolution. Fig. \ref{fig:model_archit} (c) illustrates the internal structure of the LF-DSC. It includes a pointwise convolution, an LF-DSC, and a second pointwise convolution for combining the channel features. Suppose a subview-wise 2D convolution takes an $u_i\times v_i\times x_i\times y_i\times c_i$ input tensor $T_i$ with kernel $K_j \in R^{k_j\times k_j\times c_i\times c_j}$, we can calculate the computational cost as:
\begin{equation}u_i\cdot v_i\cdot x_i\cdot y_i\cdot c_i\cdot c_j\cdot k_j\cdot k_j\label{eq:cc_conv}\end{equation}
The proposed LF-DSC can achieve similar performance but only takes:
\begin{equation}u_i\cdot v_i\cdot x_i\cdot y_i\cdot c_i(c_j + k_j^2)\label{eq:cc_dwconv}\end{equation}
Therefore, by adopting the LF-DSC, we can save the computational cost of:
\begin{equation}u_i\cdot v_i\cdot x_i\cdot y_i\cdot c_i[(c_j-1)\cdot (k_j^2-1) - 1]\label{eq:cc_reduce2d}\end{equation}
Lastly, we have two types of LF-DCS, including the stride-2 LF-DSC, which reduces the spatial dimensions and enriches the channel dimensions; and the stride-1 LF-DSC, which remains the input dimensions but applies residual shortcuts for more efficient training.

\subsection{Anglewise Separable Convolutions for LFI}
\label{subsec:ang_con}
LF-DSC is mainly designed to concentrate on spatial feature extraction. To supplement the angular quality information missed by LF-DSC, we theoretically extend LF-DSC to the angular domain of LFIs and introduce the novel concept of LF-ASC. Since LF-ASC is designed for the distinctive structure of LFIs, it makes extensive use of the extra angular information to achieve high-performance feature extraction at a low computational cost.  Fig. \ref{fig:model_archit} (b) illustrates the structure of an LF-ASC module. LF-ASC divides a one-step 4D convolution with kernel $K_j \in R^{a_j\times a_j\times k_j\times k_j\times c_i\times c_j}$ into two lower-dimensional convolutions: a horizontal $H_j \in R^{a_j\times k_j\times k_j\times c_i\times c_j}$ and a vertical $V_j \in R^{a_j\times k_j\times k_j\times c_i\times c_j}$. Hence, we succeed in reducing multiplications from one 4D convolution with $a_j^2$ to two 3D convolutions with $2a_j$ in total to achieve comparable feature extraction performance.

Suppose a standard 4D convolution takes a $u_i\times v_i\times x_i\times y_i\times c_i$ input tensor $T_i$ with kernel $K_j \in R^{a_j\times a_j\times k_j\times k_j\times c_i\times c_j}$, we can calculate the computational cost as:
\begin{equation}u_i\cdot v_i\cdot x_i\cdot y_i\cdot c_i\cdot c_j\cdot k_j^2\cdot a_j^2\label{eq:cc_4dconv}\end{equation}
The proposed LF-ASC can obtain similar performance but reduce the computational cost significantly with:
\begin{equation}u_i\cdot v_i\cdot x_i\cdot y_i\cdot c_i\cdot c_j\cdot k_j^2\cdot 2a_j\label{eq:cc_dwawconv}\end{equation}
The proposed LF-DSC and LF-ASC together can perform efficient feature extraction only at a cost of:
\begin{equation}u_i\cdot v_i\cdot x_i\cdot y_i\cdot c_i(c_j + k_j^2 + 2c_ja_jk_j^2)\label{eq:cc_dwawconv}\end{equation}
Therefore, by adopting the combination of the LF-DSC and LF-ASC instead of a 4D convolution, we can significantly reduce the computational cost by:
\begin{equation}u_i\cdot v_i\cdot x_i\cdot y_i\cdot c_i(c_ja_j^2k_j^2 - 2c_ja_jk_j^2 - k_j^2 - c_j)\label{eq:cc_reduce4d}\end{equation}

\subsection{Auxiliary Learning for NR-LFIQA}
\label{subsec:aux_learn}
To further improve the performance, we configure spatial feature estimation and angular feature estimation as two auxiliary tasks that support the primary NR-LFIQA task. The primary task shares the same hidden layers and weights with the two auxiliary tasks to take both spatial and angular quality into consideration. Nonetheless, each task has its own FC-NN to be optimised before the output.

For the labels of the auxiliary tasks, natural scene statistics (NSS) features from BRISQUE are adopted as the spatial quality features \cite{mittal2012no} due to their reliability (as discussed in Section \ref{subsec:iqa}), while the global direction distribution (GDD) features are used as the angular quality features according to the state-of-the-art NR-LFQA (as discussed in Section \ref{subsec:state_of_art}). Since BRISQUE is designed for 2D images, we take the average NSS features of all subviews as the labels for an LFI. The primary task is output with a single FC-NN while each auxiliary task uses a global average pooling to shrink the dimensions, followed by a double 256-node FC-NN to optimize the loss function.

We leverage the mean squared error (MSE) as the loss function for each task. The loss function of the primary task is formulated as:
\begin{equation}l_p = \frac{1}{N}\sum_{i=1}^{N} (y_i - \hat{y_i})^2 \label{eq:loss_qs}\end{equation} where $N$ is the total number of input LFIs, $y_i$ denotes the true quality score, and $\hat{y_i}$ represents the predicted score.

The loss function for the auxiliary task of spatial feature estimation can be formulated as:
\begin{equation}l_s = \frac{1}{D_sN}\sum_{i=1}^{N} \sum_{j=1}^{D_s}(s_{ij} - \hat{s_{ij}})^2 \label{eq:loss_sf}\end{equation} where $s_{ij}$ stands for each element from a spatial feature vector $S_i$ while $\hat{s_{ij}}$ is the corresponding predicted value. Since we adopted a $36\times 1$ NSS feature vector from BRISQUE, the dimension of the spatial feature $D_s$ is $36$.

The loss function for the auxiliary task of angular feature estimation can be formulated as:
 \begin{equation}l_a = \frac{1}{D_aN}\sum_{i=1}^{N} \sum_{j=1}^{D_a}(a_{ij} - \hat{a_{ij}})^2 \label{eq:loss_af}\end{equation} where $a_{ij}$ captures each element from an angular feature vector $A_i$ while $\hat{a_{ij}}$ is the corresponding predicted value. Since we adopted an $8\times 1$ GDD feature vector from the state-of-the-art NR-LFQA, the dimension of angular feature $D_a$ is 8.
 
The total training loss of the ALAS-DADS is defined as:
\begin{equation}L = l_p + \lambda(l_s + l_a) \label{eq:loss_sf}\end{equation} where  $\lambda$ is the balancing factor between the main and the auxiliary tasks. In related work such as \cite{yan2019naturalness, li2020personality, jaderberg2016reinforcement, kendall2018multi}, the typical values for $\lambda$ are 1, 0.1 and 0.01. In practice, we trained our framework with 1, 0.1, 0.01, and 0.001, and we chose 0.01 according to the experimental results. Additionally, we normalised the labels of the quality features for the auxiliary activities in order to scale the loss from three sources (i.e., quality score, spatial quality features, and angular quality features) to a comparable level and guarantee the effectiveness of the factor $\lambda$.

Therefore, the framework outputs a predicted quality score, the estimated spatial quality features, and the estimated angular quality features in response to an LFI. The determination of the quality scores is dependent on the labelling of the dataset on which the model is trained (see Section \ref{subsubsec:dataset}). Moreover, we use natural scene statistics (NSS) (i.e., a $36\times 1$ vector) from BRISQUE as spatial quality features and global direction distribution (GDD) (i.e., an $8\times 1$ vector) from the state-of-the-art NR-LFQA as angular quality features.

\begin{figure*}[t]
\includegraphics[width=\textwidth]{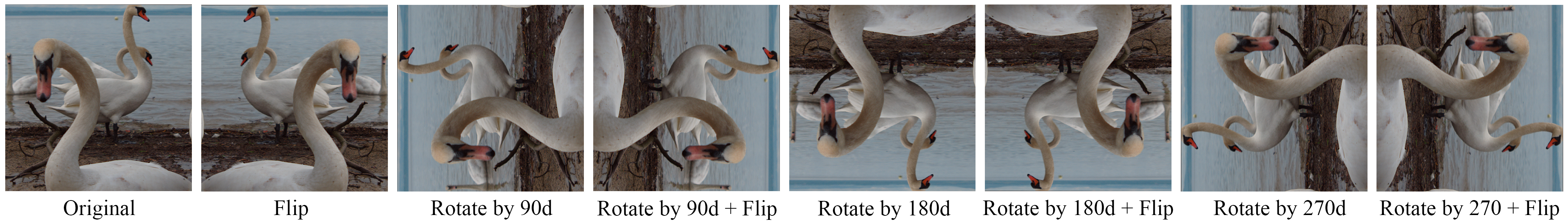}
\caption{Data Augmentation for a Sample LFI from Win5-LID \cite{shi2018perceptual}}
\label{fig:data_aug}
\end{figure*}

\subsection{Discussion on Adaptation}
\label{subsec:dis}
We believe that our proposed techniques, such as LF-DSC and LF-ASC, and the related auxiliary learning scheme, can be easily adapted to other LFI processing tasks, such as LFI super-resolution, LFI image classification, and LFI depth estimation. First, similar to how DSC is used in a variety of 2D computer vision applications \cite{zhang2019depth, ma2019lightweight, rahimian2019xceptiontime}, the LF-DSC and LF-ASC can be used as highly efficient generative LFI feature extractors for these tasks. Second, the ALAS-DADS architecture as a whole can be easily modified to perform various LFI processing tasks. For example, Lu \textit{et al.} \cite{lu2019improved} developed an LFI classification model that contains a VGGNet with an interleaved CNN refiner. One potential adaptation is to substitute the interleaved CNN in this classification model with LF-DSC and LF-ASC for more efficient feature extraction. Besides, another potential adaptation is to adopt ALAS-DADS as the framework replacing the VGGNet since VGGNet is designed for 2D images. In this instance, the primary task would be LFI classification, while the auxiliary tasks could include estimating the saliency map, which may provide useful hints to the primary task.

%\IEEEpubidadjcol

\section{Experiments}
\label{sec:experiments}
In this section, we firstly conduct an ablation study to show the effectiveness of LF-DSC and LF-ASC, and then conduct a series of experiments on mainstream LFI datasets, namely Win5-LID and SMART, for comparing the proposed metric with representative state-of-the-art metrics. Specifically, Section \ref{subsec:setup} describes the experimental setup, including the datasets (see Section \ref{subsubsec:dataset}), the data augmentation process (see Section \ref{subsubsec:data_aug}), and the performance measures (see Section \ref{subsubsec:measures}). A training scheme dedicatedly designed for LFI is introduced in Section \ref{subsec:training}. This design enables the learning of the high dimensional LFI data with limited computational resources. We also discuss the generalization of the proposed ALAS-DADS model at the end of Section \ref{subsec:training}. In Section \ref{subsec:abl_study}, we interpret the results of the ablation study to show the effectiveness and learning efficiency of LF-DSC and LF-ASC in comparison to 4D convolution. Finally, we discuss the results of the benchmarking experiments in \ref{subsec:exp_results}, which includes the performance benchmarking of the proposed metric against other metrics (see Section \ref{subsubsec:benchmarking}), performance analysis of the proposed metric in various distortion types (see Section \ref{subsubsec:anaylsis}), and some sample predictions from the proposed metric (see Section \ref{subsubsec:samples}).

\subsection{Experimental Setup}
\label{subsec:setup}
\subsubsection{Datasets}
\label{subsubsec:dataset}
We use two mainstream publicly available labelled datasets, namely Win5-LID \cite{shi2018perceptual} and SMART \cite{paudyal2017towards}, for evaluating the NR-LFIQA metrics. Since the LFIs from those two datasets have different image sizes, we trim and reshape all images into the same size of $7\times 7\times 434\times 434\times 3$, while remaining the most informative image parts according to \cite{yeung2018fast}. The Win5-LID dataset contains 220 LFIs, which are applied with 6 distortion types, including HEVC, JPEG 2000, linear interpolation (LN), nearest-neighbour (NN) interpolation, EPICNN, and USCD with five distorting levels. The quality of the 220 LFIs in Win5-LID was rated by participants on a 5-point discrete scale, under a double-stimulus continuous quality scale (DSCQS). The overall mean opinion scores (MOS) were calculated for each LFI. The SMART dataset is based on 16 original LFIs with 256 distorted LFIs obtained by 4 compression distortions, i.e., HEVC Intra, JPEG, JPEG 2000, and SSDC. The subjective ratings were gathered at the scale of the Bradley-Terry (BT) scores.

There are several points to be noted about the model output. In Win5-LID, the quality score is the mean opinion score (MOS) (which ranges from 1 to 5, with higher values indicating higher quality), whereas it is BT score (which typically ranges from -10 to 1, with higher values indicating higher quality) in SMART. The majority of spatial quality feature values are in the range (-2, 2), while the majority of angular quality feature values are in the range (-1, 1).

\subsubsection{Data Augmentation}
\label{subsubsec:data_aug}
We significantly enlarge the current datasets by data augmentation to generate more data for training, validation, and testing. To this end, according to \cite{zhang2020light}, geometric transformations are facilitated on LFIs, including rotations at different angles and vertical flipping. Although brightness, contrast, or Gaussian noise are widely used for data augmentation \cite{zhang2020light} too, those techniques are not employed because they may affect the user-perceived quality of the LFIs. Moreover, cropping is not employed either as it may impair the structural quality of the LFI. Thus, finally, we rotate the LFIs in the datasets by 90, 180, and 270 degrees, and perform vertical flipping. Finally, we increase the datasets by a factor of eight, bringing the Win5-LID to 1760 LFIs and SMART to 2048 LFIs. An example of such data augmentation is shown in Fig. \ref{fig:data_aug}.

\subsubsection{Performance Measures}
\label{subsubsec:measures}
We employ three measures to evaluate the IQA metrics' performance, including root mean square error (RMSE) \cite{dekking2005modern}, spearman rank order correlation coefficient (SROCC) \cite{zwillinger1999crc}, and Pearson linear correlation coefficient (PLCC) \cite{dekking2005modern}. The smaller the RMSE or the greater the SROCC or PLCC, the more consistent the IQA model's output is with the human perception of the LFI.

\subsection{Training and Generalization}

\begin{table}[h]
\renewcommand{\arraystretch}{1.2}
\caption{Statistics of Training}
\label{tab:tr_val_tt}
\centering
\setlength{\tabcolsep}{3pt}
\begin{tabular}{|p{0.11\textwidth}|p{0.08\textwidth}|p{0.08\textwidth}|p{0.08\textwidth}|p{0.08\textwidth}|} \hline
Dataset & Training Loss & Validation Loss & Testing Loss & Time (s)\\ \hline \hline
Win5-LID & $0.3081$& $0.3232$ & $0.3669$ & 6343 \\ \hline
SMART & $0.4560$& $0.6606$ & $0.7826$ & 14881 \\ \hline
\end{tabular}
\end{table}

\label{subsec:training}
We implemented the proposed models with Python Tensorflow, and trained them with the hardware configurations of an AMD Ryzen 7 3700X CPU, two NVIDIA RTX2070S GPUs, and 64 GB RAM. Since LFI data is high-dimensional, it is infeasible to load many LFIs simultaneously for training due to limited memory resources. Instead, we design a dedicated training scheme for efficiently processing the high dimensional LFI data. The datasets are firstly split into training and testing segments at the ratio of $0.8$ and $0.2$. We then use stochastic learning for training and validation. For each tiny training batch:
\begin{itemize}
    \item $m$ training images are randomly drawn from the training dataset without replacement.
    \item $n$ validation images are randomly drawn from the training dataset with replacement.
    \item For each epoch, the model monitors the validation performance and will early stop and switch to the next batch if:
    \begin{itemize}
        \item The performance did not improve for several epochs $p$ (i.e., patience).
        \item Or it reaches the epoch limit $l$ of a tiny training batch.
    \end{itemize}
\end{itemize}
Under this training scheme, it is possible to tune the training parameters to control the training process. For example, the epoch limit $l$ can be configured as a relatively small number (e.g., 5) to avoid overfitting. For optimization, Adam with AMS gradient \cite{reddi2019convergence} is applied, which stores ``long-term memory'' of past gradients. A new variant of the Adam algorithm is performed to solve the convergence issues and improve the performance. Since we have two GPUs, we implemented a parallel training scheme to improve the training efficiency. Due to the communication loss, Tensorflow's default distributed framework does not take advantage of multi-GPU computing. As a result, we train a model independently on two GPUs and update both with the model with the lowest validation loss in every 100 batches. Finally, as shown in TABLE \ref{tab:tr_val_tt}, it took 6343 seconds for training the model on Win5-LID and 14881 seconds on SMART.

The deep learning-based methods are often questioned with overfitting problems. To address this issue, we use residual shortcut and dropout for regularization of the CNN in addition to our data augmentation for more generalization. More importantly, we split the dataset by training and testing at the beginning. During the whole training and validation process, only the training segment is involved, while all the experimental results are based on the independent test segment without retraining. The statistics of the training process from TABLE \ref{tab:tr_val_tt} prove that the ALAS-DADS model gains an excellent trade-off between overfitting and underfitting: the training loss is slightly less than the validation loss while either of them is less than the test errors.

\begin{table}[h]
\renewcommand{\arraystretch}{1.2}
\caption{Ablation Study: 10-4D-Conv vs 10-LF-DSC vs 10-LF-ASC vs 10-LF-DSC-ASC}
\label{tab:abl_study}
\centering
\setlength{\tabcolsep}{3pt}
\begin{tabular}{|p{0.12\textwidth}|p{0.06\textwidth}|p{0.05\textwidth}|p{0.06\textwidth}|p{0.06\textwidth}|p{0.06\textwidth}|} \hline
Model & Tr. Param. & Time (s) &RMSE& SROCC& PLCC\\ \hline \hline
10-4D-Conv & 374,072 & 1587 & $2.0874$& $0.0964$ & $ 0.1140$ \\ \hline
10-LF-DSC  & 25,232 & 851 & $0.9854$& $0.1697$ & $0.1981$ \\ \hline
10-LF-ASC  & 20,222 & 690 & $1.0220$& $0.1852$ & $0.2674$ \\ \hline
10-LF-DSC-ASC  & 45,452 & 1095 & \boldmath$0.9356$& \boldmath$0.2602$ & \boldmath$0.3045$ \\ \hline
\end{tabular}
\end{table}

\subsection{Effectiveness of LF-DSC and LF-ASC}
\label{subsec:abl_study}
We performed an ablation study prior to the benchmarking experiments to demonstrate the efficacy and learning efficiency of LF-DSC and LF-ASC in comparison to 4D convolution. We developed four testing models: 10-4D-Conv, 10-LF-DSC, 10-LF-ASC, and 10-LF-DSC-ASC. Except for the backbone, the construction of the four models is identical. For example, ten four-dimensional convolution layers form the basis of 10-4D-Conv. Similarly, the 10-LF-DSC, 10-LF-ASC, and 10-LF-DSC-ASC configurations include ten LF-DSCs, ten LF-ASCs, and ten LF-DSC LF-ASC combinations, respectively. The models are then evenly trained on Win5-LID using 100 batches of 5 epochs each. Each batch contains two LFIs for training and two LFIs for validation. 
TABLE \ref{tab:abl_study} summarises the results of the ablation study. There are two things to be noted: 1) the training times in the table include data loading but not the evaluation time; and 2) the performance measures (i.e., RMSE, SROCC, and PLCC) are computed using the testing data. The number of trainable parameters in 10-4D-Conv is much larger than those in the other three models. While 10-LF-DSC-ASC includes a total of twenty layers, including ten LF-DSCs and ten LF-ASCs, it nevertheless has a much smaller number of trainable parameters and training duration than 10-4D-Conv. Additionally, despite the shorter learning period, the 10-LF-DSC and 10-LF-ASC beat the 10-4D-Conv, with the 10-LF-DSC-ASC achieving the best performance. Thus, the results of this ablation study confirms the theoretical conclusion we made in Section \ref{sec:methods} that LF-DSC and LF-ASC can attain higher NR-LFIQA performance at a reduced computational cost.

\subsection{Benchmarking Experiments}
\label{subsec:exp_results}
To validate the proposed metric's performance, we conduct an experiment comparing it to traditional metrics (i.e., PSNR, SSIM, and BRISQUE) and the state-of-the-art LFI metric (i.e., NR-LFQA). It is worth to mention that we acquire the source code of NR-LFQA from the original authors and retrain the model using identical experimental conditions. Because PSNR and SSIM are both FR-IQA metrics, we use the original undistorted images in both datasets. Apart from that, we use the average quality score of each subview for the predictions of BRISQUE.

\begin{table}[h]
\renewcommand{\arraystretch}{1.2}
\caption{Overall Results of Tested Metrics}
\setlength{\tabcolsep}{2pt}
\begin{tabular}{P{0.09\textwidth}|P{0.06\textwidth}P{0.06\textwidth}P{0.05\textwidth}|P{0.06\textwidth}P{0.06\textwidth}P{0.05\textwidth}}
\hline
\hline
&\multicolumn{3}{c}{Win5-LID}&\multicolumn{3}{c}{SMART}\\
\hline
Metrics& RMSE& SROCC & PLCC& RMSE& SROCC & PLCC \\
\hline
PSNR &  $0.8469$&  $0.6579$& $0.6577$&  $3.0367$&  $0.6996$& $0.6530$ \\
SSIM &  $1.3253$&  $0.5538$& $0.5523$&  $2.3946$&  $0.6043$& $0.6195$ \\
BRISQUE &  $0.7901$&  $0.5647$& $0.6111$&  $1.4480$&  $0.6331$& $0.7398$ \\
NR-LFQA &  $0.6421$&  $0.7346$& $0.7451$&  $1.7500$&  $0.6944$& $0.7784$ \\
\textbf{ALAS-DADS} &  \boldmath$0.3669$&  \boldmath$0.9260$& \boldmath$0.9257$&  \boldmath$0.7826$&  \boldmath$0.8540$& \boldmath$0.9344$ \\
\hline
\hline
\end{tabular}
\label{tab:BM_results}
\end{table}

\begin{table*}[h!]
\renewcommand{\arraystretch}{1.2}
\centering
\caption{RMSE and PLCC of Tested Metrics in Different Types of Distortions}
\setlength{\tabcolsep}{2pt}
\begin{tabular}{P{0.085\textwidth}|P{0.083\textwidth}P{0.083\textwidth}|P{0.083\textwidth}P{0.083\textwidth}|P{0.083\textwidth}P{0.083\textwidth}|P{0.083\textwidth}P{0.085\textwidth}|P{0.085\textwidth}P{0.085\textwidth}} \hline \hline
Distortion&\multicolumn{2}{c}{PSNR}&\multicolumn{2}{c}{SSIM}&\multicolumn{2}{c}{BRISQUE}&\multicolumn{2}{c}{NR-LFQA}&\multicolumn{2}{c}{\textbf{ALAS-DADS}}\\ \hline
\textbf{Win5-LID}&RMSE&PLCC&RMSE&PLCC&RMSE&PLCC&RMSE&PLCC&RMSE&PLCC\\ \hline
EPICNN&$1.7168$&$0.6834$&$1.0114$&$0.4374$&$0.8895$&$0.7949$&$0.8475$&$0.8131$&\boldmath$0.2002$&\boldmath$0.9895$\\
HEVC&$1.0146$&$0.5770$&$1.1218$ &$0.6354$&$0.8517$&$0.6660$&$0.6617$&$0.8307$&\boldmath$0.4402$&\boldmath$0.9281$ \\
JPEG 2000&$0.8007$&$0.6951$&$1.0953$&$0.5447$&$0.6729$&$0.7178$&$0.6306$&$0.7451$&\boldmath$0.3604$&\boldmath$0.9230$\\
LN&$0.5671$&$0.8088$&$1.6386$&$0.7039$&$0.7967$&$0.4784$&$0.7727$&$0.5609$&\boldmath$0.3263$&\boldmath$0.9298$\\
NN&$0.6070$&$0.8044$&$1.3490$&$0.7932$&$0.6804$&$0.4459$&$0.3797$&$0.8565$&\boldmath$0.3560$&\boldmath$0.8758$\\
USCD&$1.0142$&$0.5357$&$1.6746$&$0.1630$&$1.1819$&$0.4592$&$0.8802$&$0.4798$&\boldmath$0.3300$&\boldmath$0.9536$\\
\textbf{Overall} &  $0.8469$&   $0.6577$&$1.3253$& $0.5523$&$0.7901$& $0.6111$&$0.6421$& $0.7451$&  \boldmath$0.3669$& \boldmath$0.9257$\\
\hline
\textbf{SMART}&RMSE&PLCC&RMSE&PLCC&RMSE&PLCC&RMSE&PLCC&RMSE&PLCC\\ \hline
HEVC&$2.7776$&$0.7667$&$2.5174$&$0.6644$&$1.4305$&$0.8975$&$1.7701$&$0.8397$&\boldmath$0.7733$&\boldmath$0.9607$\\
JPEG&$2.2259$&$0.5868$&$2.8726$&$0.4600$&$1.3584$&$0.4173$&$1.3060$&$0.6445$&\boldmath$0.8153$&\boldmath$0.8240$ \\
JPEG 2000&$3.5570$&$0.7744$&$2.2063$&$0.7486$&$1.5690$&$0.7476$&$2.1327$&$0.8021$&\boldmath$0.7313$&\boldmath$0.9555$\\
SSDC&$3.4005$&$0.6131$&$1.8653$&$0.6554$&$1.3878$&$0.5346$&$1.6885$&$0.6946$&\boldmath$0.8079$&\boldmath$0.8899$\\
\textbf{Overall} & $3.0367$& $0.6530$&$2.3946$& $0.6195$&  $1.4480$& $0.7398$&  $1.7500$& $0.7784$&  \boldmath$0.7826$&  \boldmath$0.9344$ \\
\hline \hline
\end{tabular}
\label{tab:BM_disto_results}
\end{table*}

\subsubsection{Benchmarking Results}
\label{subsubsec:benchmarking}
TABLE \ref{tab:BM_results} shows the overall testing results of all the tested metrics. In general, ALAS-DADS significantly outperforms the others in either Win5-LID or SMART. In Win5-LID specifically, ALAS-DADS gains $0.2752$ RMSE reduction as well as $0.1914$ and $0.1806$ improvements on SROCC and PLCC respectively, compared with the state-or-art NR-LFQA model. Moreover, ALAS-DADS gains a more significant performance boost in SMART where it achieves $0.6654$ less RMSE than BRISQUE (the second-best) and increases the SROCC and PLCC by $0.1596$ and $0.1560$, respectively, compared to the state-of-the-art NR-LFQA.

The performance of the tested metrics in different distortion types are presented in TABLE \ref{tab:BM_disto_results}. One can see that ALAS-DADS gains the lowest RMSE for all the distortion types in both datasets. In the most challenging distortion types of Win5-LID, e.g., EPICNN and USCD, the ALAS-DADS achieves the stunning $0.2002$ and $0.3300$ of RMSE, respectively, while the other metrics obtain more than $0.8400$ in both distortion types. In addition, it is noticed that the ALAS-DADS can also substantially shrink the RMSE for SMART in each distortion type compared to other metrics.

\begin{figure*}[h!]
\includegraphics[width=\textwidth]{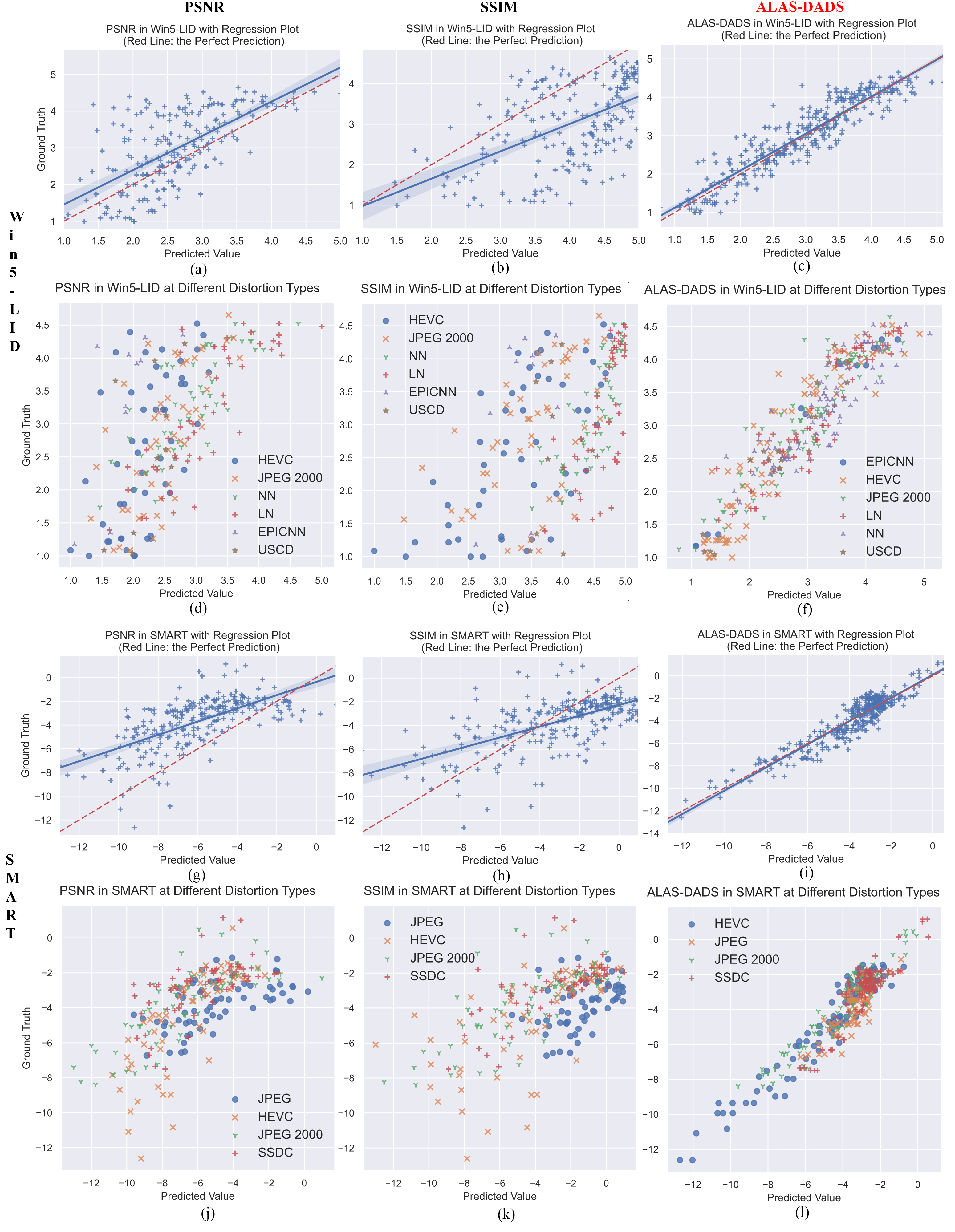}
\caption{Comparison of Regression and Distortion Scatter Plots Between ALAS-DADS and Other FR-IQA Metrics}
\label{fig:bench_plots_1}
\end{figure*}

\begin{figure*}[h!]
\includegraphics[width=\textwidth]{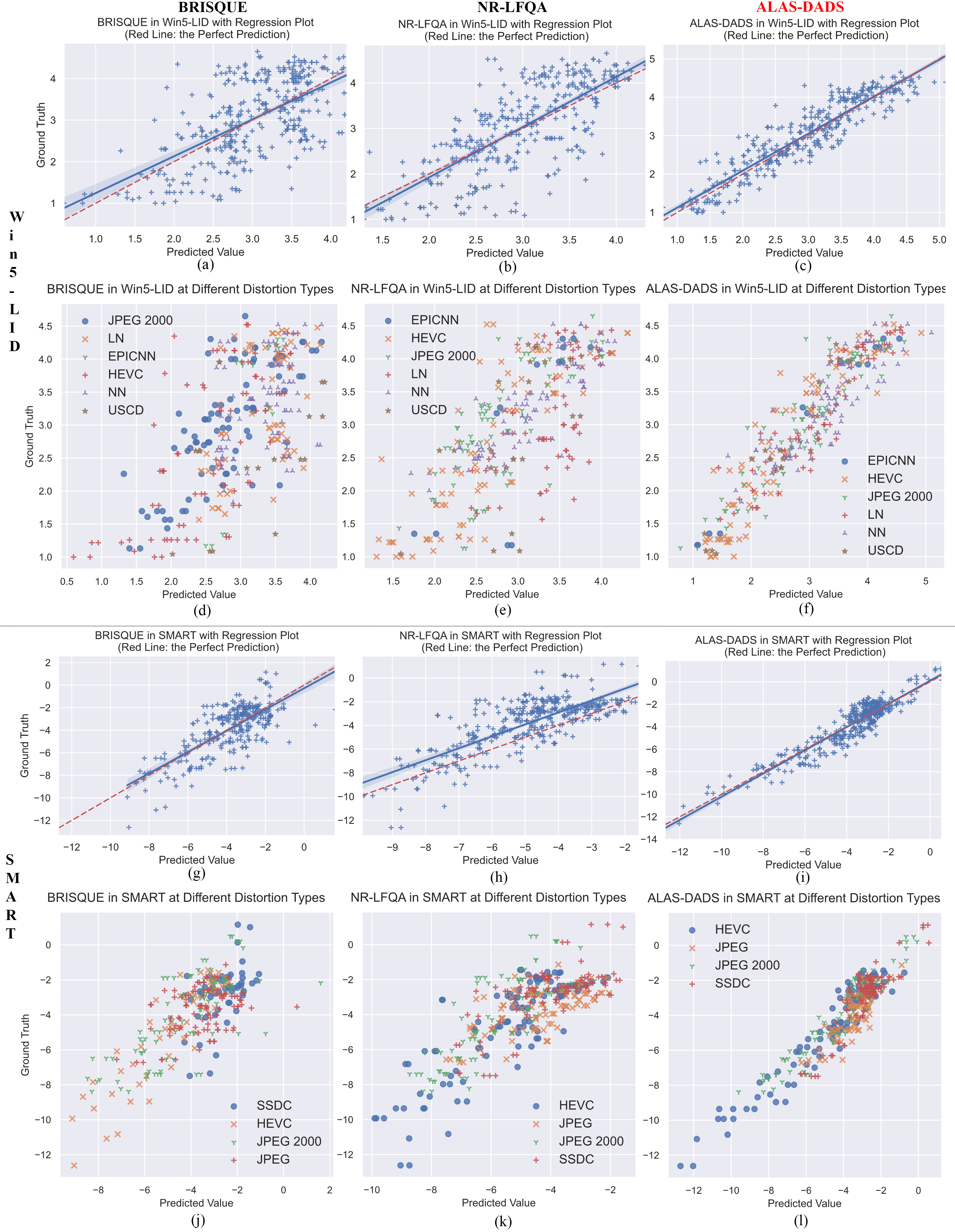}
\caption{Comparison of Regression and Distortion Scatter Plots Between ALAS-DADS and Other NR-IQA Metrics}
\label{fig:bench_plots_2}
\end{figure*}

Fig. \ref{fig:bench_plots_1} and \ref{fig:bench_plots_2} visualise the benchmarking results with the FR-IQAs as well as the NR-IQAs in each dataset. Both figures can be read from left to right to notice the significant improvement the proposed model achieved. In the comparison between regression scatter plots, ALAS-DADS's predicted quality scores are more centralized and closer to the red line (i.e., the perfect prediction) than those of the other metrics on both datasets. The blue line (i.e., linear regression) almost overlaps the red line, which implies that ALAS-DADS predicts not only more unbiased but also with lower variances compared with other metrics. From the scatter plots of each distortion type, we can draw the same conclusion: in each type of distortion, ALAS-DADS can consistently achieve more accurate quality scores than the others.

To sum up, ALAS-DADS gains $0.2752$ (i.e., 42.86\%) RMSE reduction and $0.1914$ and $0.1806$ improvements on SROCC and PLCC, respectively, compared with the second-best state-of-the-art NR-LFQA model in Win5-LID. Moreover, ALAS-DADS gains a more prominent performance boost in SMART where it achieves $0.6654$ (i.e., 45.95\%) less RMSE than the second-best BRISQUE, and increases the SROCC and PLCC by $0.1596$ (i.e., 22.98\%) and $0.1560$ (i.e., 20.04\%) respectively, compared to the state-of-the-art NR-LFQA. In the most challenge distortion types (i.e., EPICNN and USCD in Win5-LID), ALAS-DADS achieves the stunning $0.2002$ (i.e., 76.38\% better than the state-of-the-art) and $0.3300$ (i.e., 62.51\% better than the state-of-the-art) of RMSE, respectively, while the other metrics obtain more than $0.8400$ in both distortion types.

\subsubsection{Performance Analysis}
\label{subsubsec:anaylsis}
\begin{figure*}[h!]
\includegraphics[width=\linewidth]{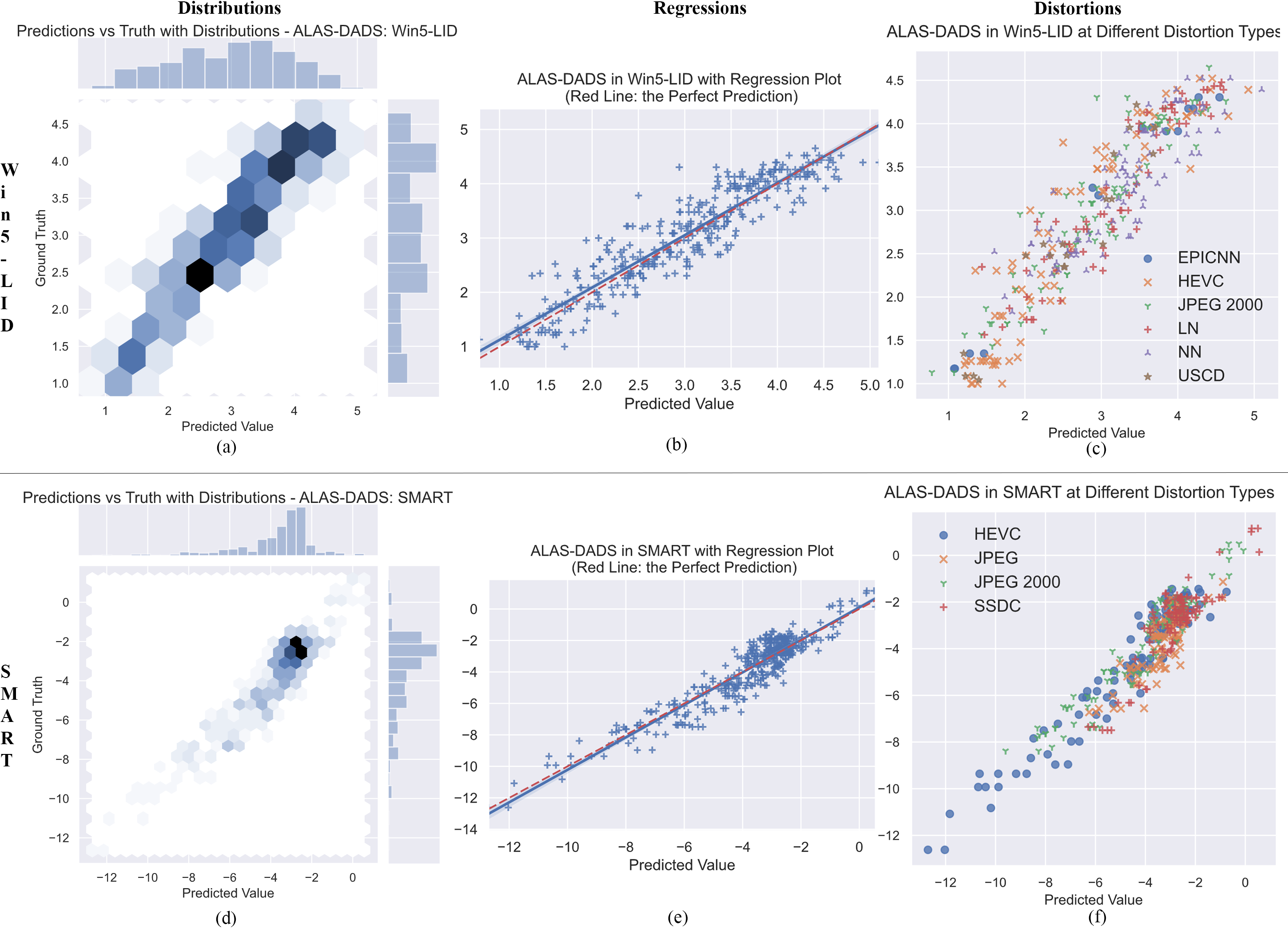}
\caption{Scatter Plots Showing Predicted Values of ALAS-DADS vs Ground Truth Values}
\label{fig:ADresults}
\end{figure*}

To further visualize the performance yielded by the proposed model, Fig. \ref{fig:ADresults} shows the scatter plots of the predicted quality scores versus ground truth. In Fig. \ref{fig:ADresults} (a) and (d), the x-axis demonstrates distributions of the quality score predictions and ground truth where colored hexagons indicate the density of points in that area. Fig. \ref{fig:ADresults} (a) is based on the testing results in Win5-LID, where the predicted values are in a normal distribution while the ground truth values have a more flat shape. Fig. \ref{fig:ADresults} (d) is based on SMART where either predicted values or ground truth are like a normal distribution.

The blue lines in Fig. \ref{fig:ADresults} (b) and (e) are the linear regression of all the prediction points, and the red dashed line is the perfect prediction line with the slop of 1, which is used as the anchoring line. In either Fig. \ref{fig:ADresults} (b) or (e), it can be seen that the linear regression of the predicted values output from ALAS-DADS is very close to the perfect prediction line. The predictions in Fig. \ref{fig:ADresults} (e) are more concentrated on the ideal prediction line than those in Fig. 5 (b). As shown in Fig. 5 (a) and (d), the distribution of ground truth of SMART is more predictable (i.e., closer to a normal distribution) than that of Win5-LID, which may bring the lower variance in the ALAS-DADS’s predictions for SMART.

\begin{table}[h!]
\renewcommand{\arraystretch}{1.2}
\centering
\caption{Performance Details of ALAS-DADS in Different Distortion Types}
\setlength{\tabcolsep}{2pt}
\begin{tabular}{P{0.2\linewidth}|P{0.2\linewidth}|P{0.17\linewidth}P{0.17\linewidth}P{0.17\linewidth}}
\hline \hline
Dataset&Distortion&RMSE&SROCC&PLCC\\ \hline
\multirow{7}{*}{Win5-LID}&EPICNN&\boldmath$0.2002$&\boldmath$0.9536$&\boldmath$0.9895$\\
&HEVC&$0.4402$&$0.9369$&$0.9281$ \\
&JPEG 2000&$0.3604$&$0.9143$&$0.9230$\\
&LN&$0.3263$&$0.9363$&$0.9298$\\
&NN&$0.3560$&$0.8785$ &$0.8758$\\
&USCD&$0.3300$&$0.8937$&$0.9536$\\ \hhline{~|----}
&\textbf{Overall}&$0.3669$&$0.9260$&$0.9257$\\ \hline\hline
\multirow{5}{*}{SMART}&HEVC&$0.7733$&$0.9194$&\boldmath$0.9607$\\
&JPEG&$0.8153$&$0.7404$&$0.8240$\\
&JPEG 2000&\boldmath$0.7313$&\boldmath$0.9288$&$0.9555$\\
&SSDC&$0.8079$&$0.6842$&$0.8899$\\ \hhline{~|----}
&\textbf{Overall}&$0.7826$&  $0.8540$& $0.9344$\\ \hline \hline
\end{tabular}
\label{tab:AD_disto_results}
\end{table}

Fig. \ref{fig:ADresults} (c) and (f) show the scatter plots of the predicted values and the ground truth in each type of distortions on each dataset, showing that ALAS-DADS predicts very close to the ground truth in each distortion type (performance details are shown in TABLE \ref{tab:AD_disto_results}). In dataset Win5-LID, it turns out that ALAS-DADS performs well at predicting the quality score of EPICNN distorted LFIs, achieving the smallest RMSE of $0.2002$ and the best PLCC of $0.9895$ (i.e., extremely close to the perfect $1$), while it obtains the least SROCC and PLCC in LFIs distorted by nearest-neighbour interpolation. In the SMART dataset, ALAS-DADS tends to predict more accurately for the HEVC and JPEG 2000 distorted LFIs than the JPEG and SSDC distorted LFIs. In the above results, we noticed the performance variations in predictions for different distortion types. We believe this might be affected by the ALAS-DADS's adoption of the spatial and angular quality features (e.g., NSS from BRISQUE as spatial quality feature and GDD from NR-LFQA as angular quality feature), which act as hints for auxiliary learning. For example, both NR-LFQA with NSS and BRISQUE with GDD perform well in JPEG 2000, which could help the ALAS-DADS to obtain a higher accuracy for JPEG 2000, too.

\subsubsection{Sample Predictions}
\label{subsubsec:samples}

\begin{figure}[h!]
\includegraphics[width=\linewidth]{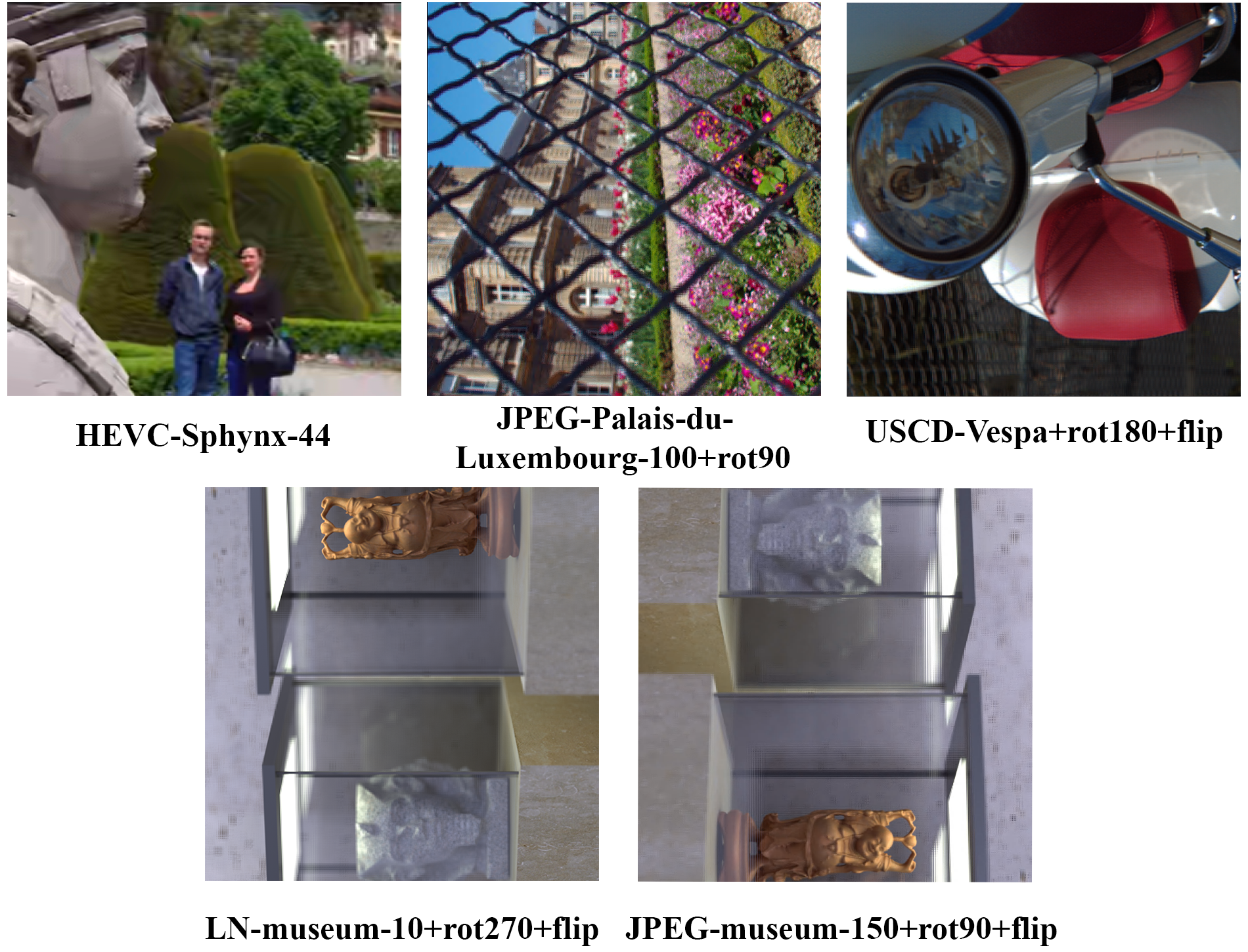}
\caption{Sample LFIs from Win5-LID}
\label{fig:example_img_w}
\end{figure}
\begin{figure}[h!]
\includegraphics[width=\linewidth]{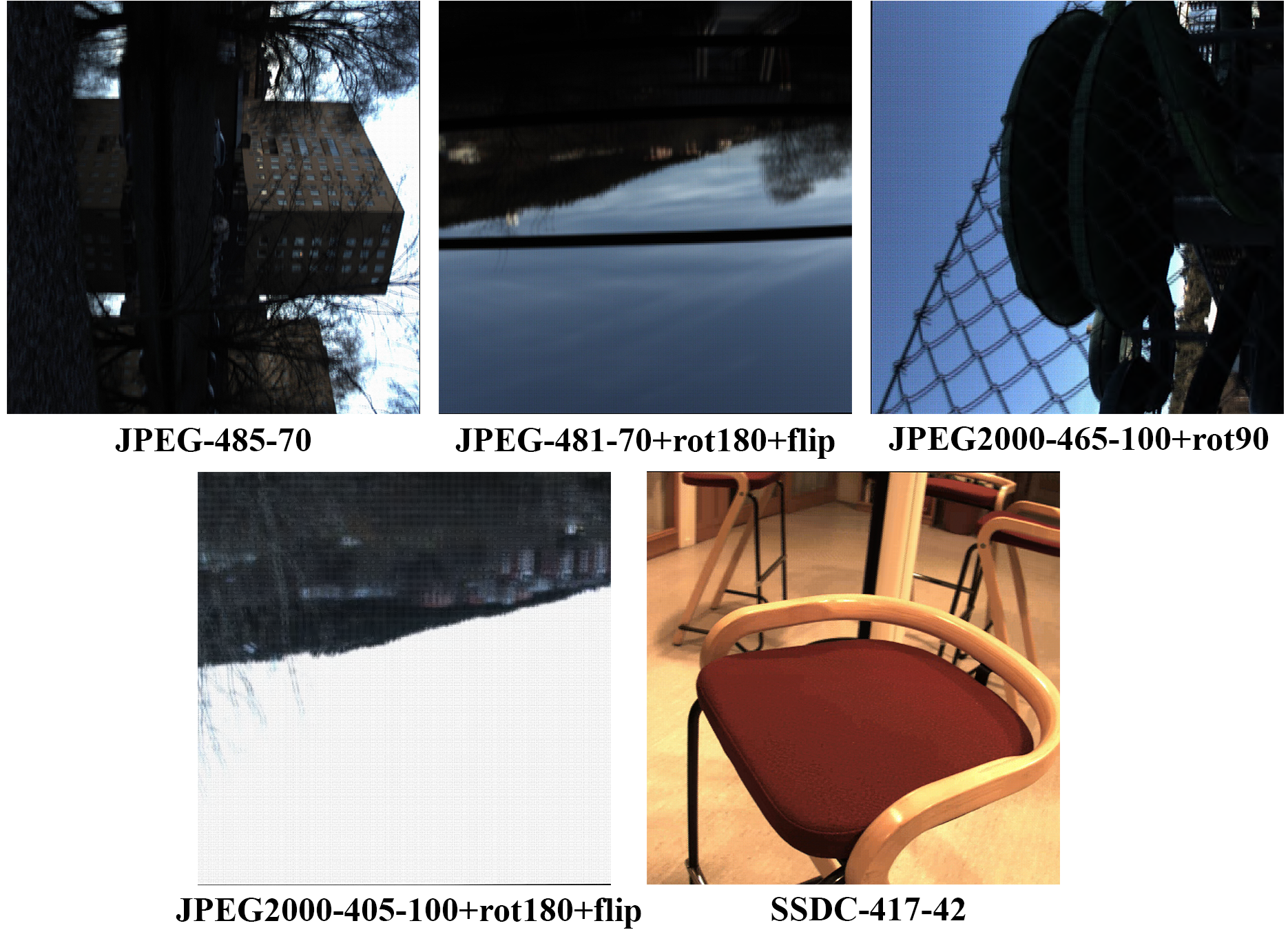}
\caption{Sample LFIs from SMART}
\label{fig:example_img_s}
\end{figure}

\begin{table}[h!]
\renewcommand{\arraystretch}{1.2}
\centering
\caption{Example Image Results for Fig. \ref{fig:example_img_w} and Fig. \ref{fig:example_img_s}}
\setlength{\tabcolsep}{2pt}
\begin{tabular}{P{0.25\linewidth}|P{0.14\linewidth}|P{0.19\linewidth}P{0.15\linewidth}P{0.18\linewidth}} \hline \hline
Image&True&Method&Pred&$|$Pred-True$|$\\ \hline
\multicolumn{5}{c}{Win5-LID}\\ \hline
\multirow{3}{*}{\shortstack{HEVC-Sphynx-44}}&\multirow{3}{*}{$1.2174$}&BRISQUE&$1.2835$&$0.0661$\\
&&NR-LFQA&$1.5041$&$0.2867$\\
&&\textbf{ALAS-DADS}&$1.2125$&\boldmath$0.0049$\\
\hline
\multirow{3}{*}{\shortstack{JPEG2000-Palais-du-\\Luxembourg-\\100+rot90}}&\multirow{3}{*}{$2.9130$}&BRISQUE&$3.0901$&$0.1770$\\
&&NR-LFQA&$3.1556$&$0.2425$\\
&&\textbf{ALAS-DADS}&$2.9191$&\boldmath$0.0060$\\
\hline
\multirow{3}{*}{\shortstack{USCD-Vespa\\+rot180+flip}}&\multirow{3}{*}{$3.1304$}&BRISQUE&$3.9864$&$0.8559$\\
&&NR-LFQA&$3.7500$&$0.6196$\\
&&\textbf{ALAS-DADS}&$3.1367$&\boldmath$0.0062$\\
\hline
\multirow{3}{*}{\shortstack{LN-museum-10\\+rot270+flip}}&\multirow{3}{*}{$4.3913$}&BRISQUE&$3.4670$&$0.9243$\\
&&NR-LFQA&$2.6758$&$0.1503$\\
&&\textbf{ALAS-DADS}&$2.8190$&\boldmath$0.0070$\\
\hline
\multirow{3}{*}{\shortstack{JPEG2000-museum-\\150+rot90+flip}}&\multirow{3}{*}{$2.8261$}&BRISQUE&$3.4670$&$0.9243$\\
&&NR-LFQA&$3.4795$&$0.9118$\\
&&\textbf{ALAS-DADS}&$4.3850$&\boldmath$0.0063$\\
\hline
 
\multicolumn{5}{c}{SMART}\\ \hline
\multirow{3}{*}{\shortstack{JPEG-485-70\\+rot90+flip}}&\multirow{3}{*}{$-2.9399$}&BRISQUE&$-4.1971$&$0.7319$\\
&&NR-LFQA&$-2.4821$&$0.4579$\\
&&\textbf{ALAS-DADS}&$-3.4729$&\boldmath$0.0078$\\
\hline
\multirow{3}{*}{\shortstack{JPEG-481-70\\+rot180+flip}}&\multirow{3}{*}{$-3.4652$}&BRISQUE&$-4.1971$&$0.7319$\\
&&NR-LFQA&$-4.5941$&$1.1290$\\
&&\textbf{ALAS-DADS}&$-2.9453$&\boldmath$0.0054$\\
\hline
\multirow{3}{*}{\shortstack{JPEG2000-465-\\100+rot90}}&\multirow{3}{*}{$3.1304$}&BRISQUE&$-6.1960$&$0.8239$\\
&&NR-LFQA&$-5.5500$&$0.1779$\\
&&\textbf{ALAS-DADS}&$-6.1495$&\boldmath$0.0118$\\
\hline
\multirow{3}{*}{\shortstack{JPEG2000-405-\\100+rot180+flip}}&\multirow{3}{*}{$4.3913$}&BRISQUE&$-6.1064$&$0.0549$\\
&&NR-LFQA&$-6.6884$&$0.5271$\\
&&\textbf{ALAS-DADS}&$-5.3621$&\boldmath$0.0100$\\
\hline
\multirow{3}{*}{\shortstack{SSDC-417-42}}&\multirow{3}{*}{$-2.3157$}&BRISQUE&$-1.9397$&$0.3759$\\
&&NR-LFQA&$-1.7745$&$0.5411$\\
&&\textbf{ALAS-DADS}&$-2.3038$&\boldmath$0.0118$\\
\hline
\hline
\end{tabular}
\label{tab:exp_results}
\end{table}

From each dataset, we choose five sample LFIs for demonstrating the prediction results. The results are shown in TABLE \ref{tab:exp_results} and the associated images are displayed in Fig. \ref{fig:example_img_w} and \ref{fig:example_img_s}. For the samples of Win5-LID, the differences between the predicted values and the ground truth values are all smaller than $0.01$, which is significantly less than that of the other NF-LFIQA metrics. For SMART, the ALAS-DADS likewise can achieve remarkable performance with errors averaging about $0.01$.

\section{Conclusion}
\label{sec:conclusion}
In this paper, we propose an improved NR-LFIQA metric called ALAS-DADS. To the best of our knowledge, this work is the first exploration of deep auxiliary learning with spatial-angular hints on NR-LFIQA. Particularly, the proposed ALAS-DADS equips LF-DSC and LF-ASC, which both are highly effective and low-complexity feature extractors. To utilize the spatial and angular hints obtained from these feature extractors, we design an auxiliary tasking scheme consisting of spatial and angular feature estimation sub-tasks to assist the primary task of quality assessment for more accurate IQA results. The experimental results show that ALAS-DADS significantly outperforms both mainstream FR-IQA metrics and the state-of-the-art NR-LFIQA methods. Overall, it achieves more than 40\% performance boost and showed robustness in various distortion types. In some categories, it gains significant improvement of more than 60\%.

However, we are aware that our work is limited in the following aspects. First, our exploration of feature labels in the auxiliary tasks may not be exhaustive. In our future work, we will attempt updating the labels of auxiliary tasks with more advanced features, e.g., updating either BRISQUE features or GDD features for further improvement of the model. Second, our investigation on the auxiliary tasks may not be thorough. In the future, we will attempt introducing more auxiliary tasks, e.g., a new learning branch (sub-task) estimating the saliency map \cite{yang2018saliency} to assist the primary IQA task. Third, due to the limitation of current public LFI datasets, we have not evaluated ALAS-DADS's performance with some distortion types, such as JPEG Pleno compression \cite{astola2020jpeg}. In our future work, we plan to establish a new LFI dataset or augment the current LFI datasets covering more advanced compression techniques including JPEG Pleno, conduct subjective experiments, and evaluate our improved model accordingly.

Despite these limitations, we believe that our work contributes to the advancement of NR-LFIQA and offers theoretical insights for a wider spectrum of LFI research. We believe that the novel concepts introduced in this study, such as LF-DSC and LF-ASC, as well as the associated auxiliary learning scheme, can be readily extended to other LFI processing tasks, such as LFI super-resolution, LFI image classification, and LFI depth estimation. For example, the LF-DSC and LF-ASC can be employed as highly efficient generative LFI feature extractors for these tasks. Besides, the ALAS-DADS architecture as a whole can also be readily adapted to better perform the above LFI processing tasks.

\ifCLASSOPTIONcaptionsoff
  \newpage
\fi

\bibliographystyle{IEEEtran}

\bibliography{main}

\end{document}